\documentclass{ptephy_v1}

\usepackage{graphicx}
\graphicspath{{./}}
\usepackage{subfig}
\usepackage[separate-uncertainty=true,multi-part-units=single]{siunitx}

\begin{document}

\title{Super resolution plasmonic imaging microscopy for submicron tracking emulsion detector}


\author{Atsuhiro Umemoto}
\affil{Graduate School of Science, Nagoya University, Furo-cho, Chikusa-ku, Nagoya,
464-8602, Japan\email{umemoto@flab.phys.nagoya-u.ac.jp}}

\author{Tatsuhiro Naka}
\affil{Kobayashi-Maskawa Institute for the Origin of Particles and the Universe, Center
for Experimental Studies, Nagoya University, Furou-cho, Chigusa-ku, Nagoya}

\author{Andrey Alexandrov}
\affil{INFN Sezione di Napoli, Napoli I-80125, Italy}

\author{Masahiro Yoshimoto}
\affil{Physics Department, Gifu University, Gifu 501-1193, Japan}


\begin{abstract}
~~NIT is a super fine-grained nuclear emulsion which has a detection capability for ionizing particle with nanometric resolution and record the track by a line of silver nanograin with a various of shape and size.
The particle tracks need to be read out by some microscopic techniques, here we focused on the optical response of silver nanograins to realize the readout method beyond the diffraction limit.
In this paper, super-resolution plasmonic imaging microscopy (SPRIM), which utilized the polarization-dependent optical response due to Localized Surface Plasmon Resonance (LSPR) was developed.
The spatial resolution of SPRIM to identify the nanograin position was achieved 5 nm.
We showed that the SPRIM clearly discriminated the 100 keV carbon ions
tracks with the mean range of 270 nm from single nanograins with the
diameter of 60 nm.The recognition efficiency of 100 keV carbon track was 49 $\%$ and the
angular resolution was 17$^\circ$.
\end{abstract}

\subjectindex{H20, H21, C44}

\maketitle

\section{Introduction}

~~The spatial resolution of a particle tracking detector is one of the most important parameters with respect to determining the kinetic characteristics and the discovery of new particles. Nuclear emulsion is a solid tracking detector with the highest spatial resolution. A super fine - grained nuclear emulsion, Nano Imaging Tracker (NIT), which was developed in recent years, has a measurement capability for particle tracking with a size less than one micrometer~\cite{NIT}. This performance is attributable to the use of very fine silver halide crystals which work as sensors to record a charged particle track. 
The number of crystals now approaches 10000 crystals/$\mathrm{\mu m^3}$ with a diameter of  44 nm, compared to 100 crystals /$\mathrm{\mu m^3}$ with a diameter of 200 nm in conventional nuclear emulsions~\cite{OPERA}. In NIT device, the average distance between the crystals is 71 nm~\cite{NIT} and this corresponds to the intrinsic tracking resolution. Therefore, NIT can potentially facilitate new opportunities in low energy physics (e.g. directional dark matter search~\cite{NEWS})and the vertex reconstruction with a large number of tracks in high energy physics. 
The particle track is recorded as a line of silver nanograins originating from the silver halide crystals via image development process. It is subsequently identified using an optical microscopy system which is utilized for large-area automatic analysis. As such, the typical spatial resolution of the optical microscope is approximately 200 nm (e.g. $\lambda$=450 nm, NA=1.4) defined based on the Rayleigh criterion however, the intrinsic tracking resolution of NIT is better. 
This means that microscopy below the diffraction limit (i.e., super-resolution) is essential in achieving the best performance in NIT device.
Numerous approaches for super-resolution have been developed so far, such as stimulated emission depletion (STED) microscopy~\cite{Hell}, spontaneous emission and photoactivated localization microscopy (PALM) / stochastic optical reconstruction microscopy (STORM)~\cite{Betzig}. They are based on the optical properties of fluorescence molecules for observing a single molecule in a large numbers, and nanostructures less than the optical resolution can be resolved by combining the data for induvial molecules ~\cite{Hell2}\cite{Cell}. As a result, the optical microscopic spatial resolution is improved to 10 nm.
Recently another super-resolution has been achieved using the optical property of metallic nanoparticles called “localized surface plasmon resonance (LSPR)”~\cite{CPL}.
LSPR is also observed from a track consisting of silver nanograins in NIT and it is therefore expected to realize super-resolution readout. This is a new technique that combines the principles of particle detection with that of physical metallurgy and it is motivated to obtain track information beyond the optical resolution limit. In this paper, we present the idea and design for a super-resolution tracking method.

\section{Plasmonic super-resolution imaging}
\subsection{LSPR in NIT}

\begin{figure}[htbp]
 \begin{center}
  \includegraphics[width=100mm]{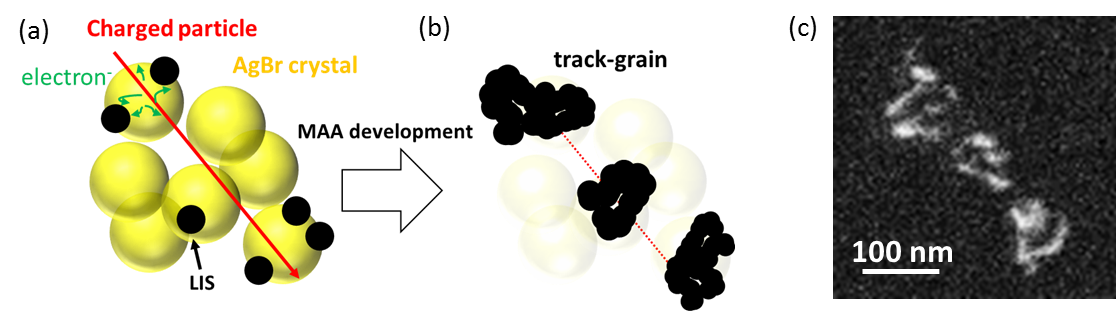}
  \caption{Schematic of the detection m
echanism of NIT and SEM image of track-grains detected by NIT. (a) is a illustration of the track detection mechanism and (b) is a illustration of track-grains after MAA development. (c) is a SEM image of track-grains after MAA development due to carbon ion with 100 keV kinetic energy. The limited depth of view in the SEM is approximately 20 nm, therefore, only the upper part of the structure can be seen.}
 \label{fig:NITdetect}
 \end{center}
\end{figure}

 ~~NIT is a super fine-grained emulsion that is developed for detecting sub-micron charged particle track with directionality. NIT consists of AgBr crystals dispersed in gelatin binder and the crystals operate as sensors to record the track of a charged particle. Schematics of the detection mechanism of NIT are shown in Fig.~\ref{fig:NITdetect}(a) (b) and a Scanning Electron Microscope (SEM) image of a 100 keV carbon ion track detected by NIT is shown in Fig.~\ref{fig:NITdetect}(c). \\

The excited electrons induced by the charged particle are captured in electron traps in the AgBr crystal. Then silver cores, so called “latent image specks” (LIS) are produced by the reaction with between electrons and interstitial silver ions. They eventually grow to a visible size (typically approximately 100 nm) under optical microscope via chemical development processes. Here after, LIS after image development is called “track-grain”. The number or size of the LIS strongly depends on the ionization (i.e., dE/dx) of the incoming charged particles. Therefore, the structure of the track-grains depends on the intensity of dE/dx~\cite{naka}. And the utilized image developing process  also affects the track-grain structure. For example, standard chemical developers (e.g. MAA developer\cite{PSA}) result in complicated filament shapes like those shown in Fig.~\ref{fig:NITdetect}(c).

\begin{figure}[htbp]
 \begin{center}
  \includegraphics[width=80mm]{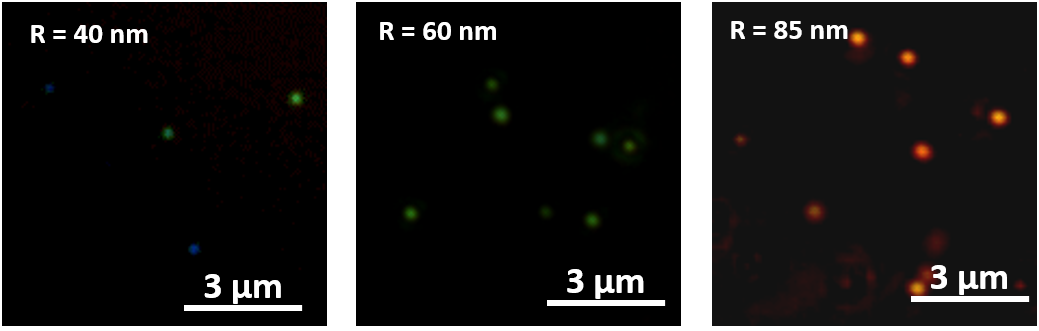}
  \caption{Color images of silver spherical nanoparticles dispersed in gelatin.
The diameter of each nanoparticle is 40 nm, 60 nm and 85 nm from left to right.
The sensitivity of the CCD sensor was not calibrated and the quantum efficiency of blue was the lowest in RGB.}
 \label{fig:Agnano}
 \end{center}
\end{figure}

In order to observe these track-grains via optical microscopy, we focused on LSPR. LSPR is a resonantly enhanced optical response of metallic nanoparticles and it can be observed when the incoming wavelength is large in comparison with the nanoparticles~\cite{OP}\cite{SmallP}. The resonance wavelength is determined by the relative dielectric constant of the nanoparticles that strongly depends on the size, shape, type of atomic element, and the surrounding medium. In this paper, an optical microscope with an epi-illumination system described in section 3-1 was used to observe the LSPR response.
Initially, we checked the LSPR of the silver spherical nanoparticles dispersed in gelatin and coated on polystyrene film which sample structure was made to become similar to that of NIT device. 
In general, the LSPR of silver nanoparticle occurs in the visible region of the spectrum~\cite{OP}. Fig.~\ref{fig:Agnano} shows the epi-illuminated optical images of the silver spherical nanoparticles of sizes  40 $\pm$ 4, 60 $\pm$ 8, and 85 $\pm$ 10 nm acquired with a color camera (SENTECH STC-CL33A) using a halogen lamp illumination source.
Each resonant wavelength appears in blue, green, and red, respectively, and it shows the relation that the larger nanoparticle has the reflection with the longer wavelength~\cite{OP}.\\

\begin{figure}[htbp]
 \begin{center}
  \includegraphics[width=115mm]{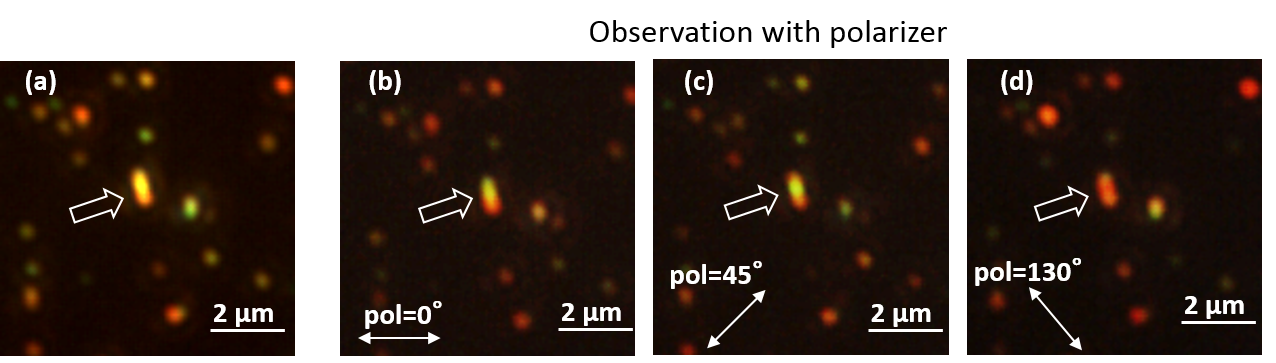}
  \caption{Optical color image of 100 keV carbon track detected by NIT. 
(a) is acquired using an epi-illuminated microscope with a color camera and (b),(c), and (d) are acquired using linear polarizer inserted before the camera. The polarizer was rotated and each rotated angle is indicated in images. The angle of image (b) is 0$^\circ$.}
 \label{fig:C100C}
 \end{center}
\end{figure}
On the other hand, various colors of reflection were observed from the track-grains due to the carbon ions. Fig.~\ref{fig:C100C} is an optical image of the 100 keV carbon track with the same optical setup used as in Fig.~\ref{fig:Agnano}. The track-grains of the 100 keV carbon ions are expected to be formed of a variety of structures and size based on the track detection mechanism shown as Fig.~\ref{fig:NITdetect}. Therefore, the resonance wavelength of each track-grain is represented by its own shape and size. Furthermore, if nanoparticles have an anisotropic shape, the resonance condition has polarization dependency \cite{tamaru}. In Fig.~\ref{fig:C100C}, images (b), (c), and (d) have the same view as the image of Fig.~\ref{fig:C100C} (a), except that they were acquired using a linear polarizer installed before the camera. Each polarization axis of Fig.~\ref{fig:C100C} (b), (c), and (d) correspond to 0$^\circ$, 45$^\circ$, 130$^\circ$ relative to the transverse direction, respectively. The track color changed according to the rotation of the polarization axis, and drastic color changes were exhibited, especially for the tracks near the center. These color changes are due to the LSPR associated with the track-grain, i.e., track-grains are present at the locations where the color changes. This provides track information beyond the optical resolution, for example, it is possible to estimate the number of track-grains (we can estimate that there are more than two track-grains in the central track marked with the arrow).  
Elliptical fitting is a standard method for detecting a submicron track direction \cite{kimura}, but tracking beyond optical limitation becomes possible due to the LSPR response. In the next section, the idea of super-resolution using LSPR is described.
\subsection{Super-resolution plasmonic imaging}
 
\begin{figure}[htbp]
 \begin{center}
   \includegraphics[width=55mm]{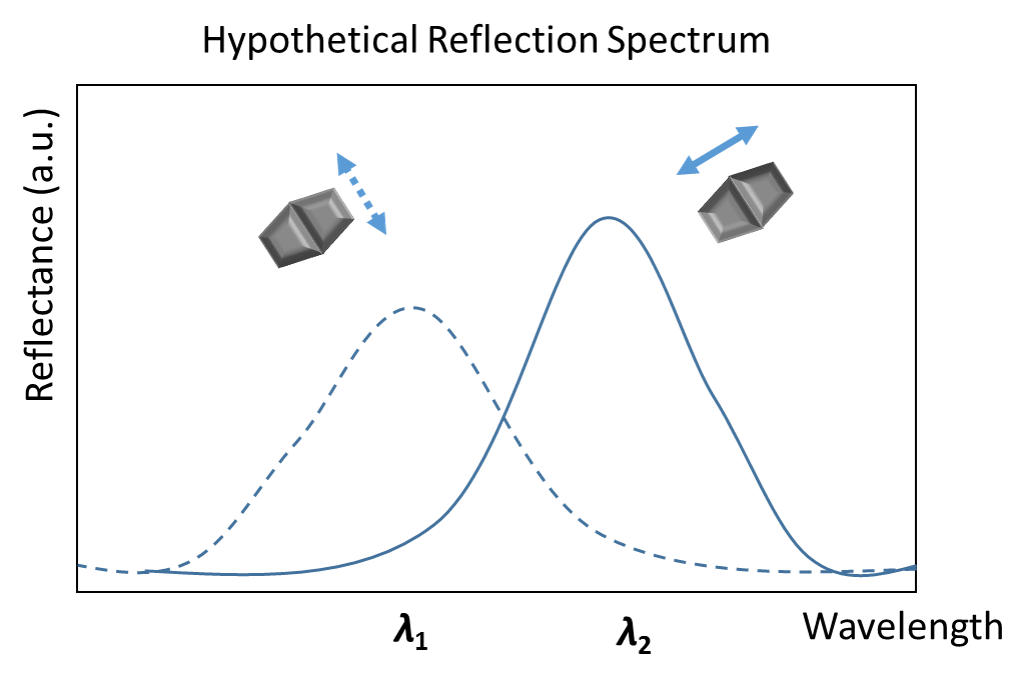}
  \caption{Hypothetical reflection spectrum of nanorods with respect to linearly polarized light. The spectrum identified with the dashed line represents the case when the polarization is in the same direction as the minor axis. The spectrum identified by the solid represents the case when the polarization is in the same direction as the major axis.}
  \label{fig:rodspec}
 \end{center}
\end{figure}

\begin{figure}[htbp]
 \begin{center}
   \includegraphics[width=75mm]{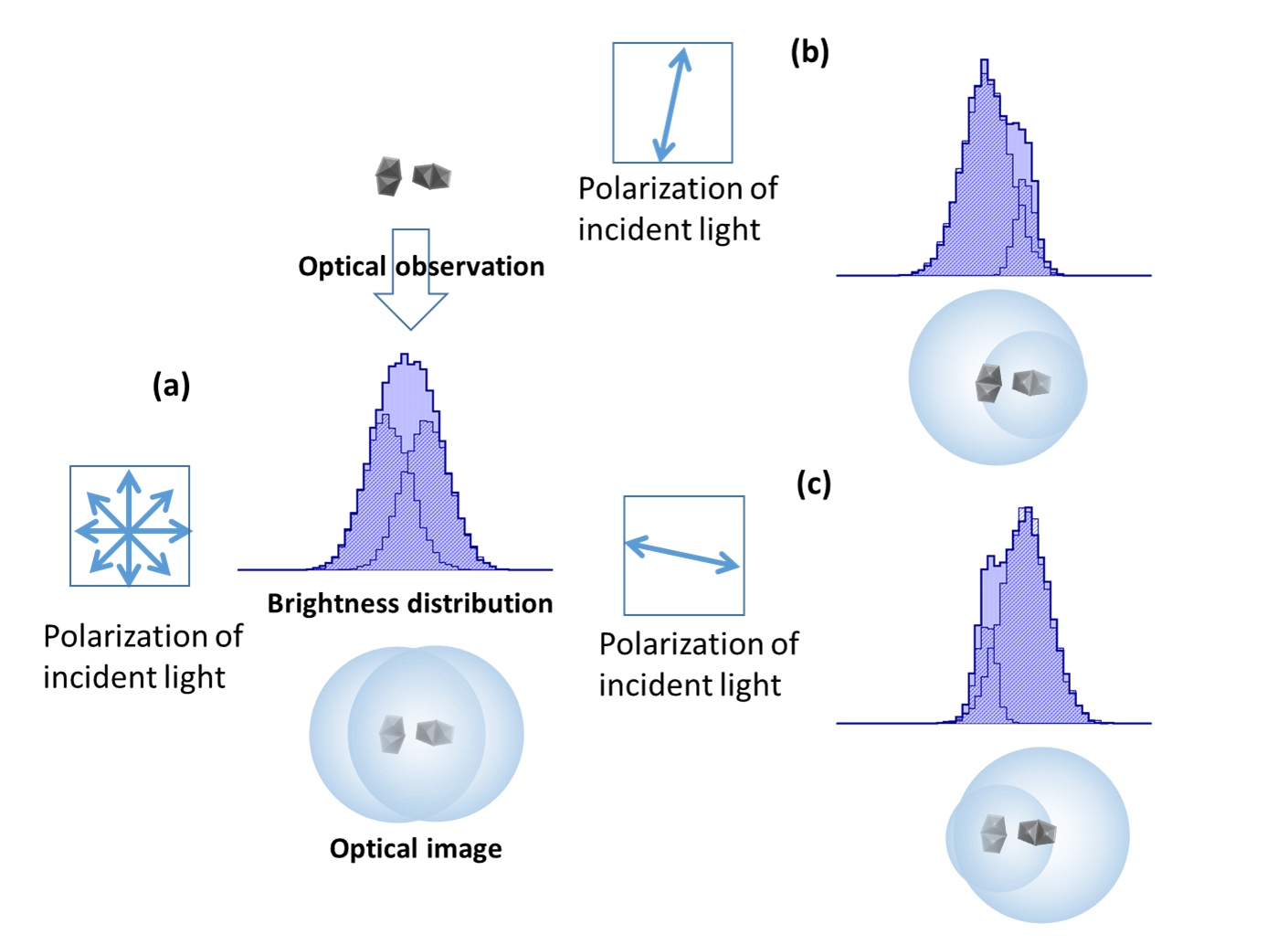}
 \caption{Schematic of the optical response of a collection of silver nanorods. 
(a) There is no difference in the optical response for the two nanorods with non-polarized illumination. The optical image appears to be one cluster due to diffraction. (b) When $\lambda_2$ is used and the linearly polarized light with a polarization axis in the same direction as the major direction of the left side of the nanorods. The stronger reflection will occur at the left side and the optical image biased to the left side. (c) The case of using $\lambda_2$ with linear polarized light along an axis in the same direction as the major direction of the right side of the nanorods. A stronger reflection will occur at the right side and this is represented in the optical image.} 
 \label{fig:spectrum}
 \end{center}
\end{figure}

~~The idea of super-resolution analysis for NIT track detection is described shortly and the approach is named as “Super-Resolution Plasmonic Imaging Microscopy (SRPIM)”.
We introduce the method of SRPIM using a monochromatic camera and monochromatic light to obtain a better LSPR response. Since a color camera has one red, blue and two green sensors per group of four pixels, the effective resolution at a given wavelength and the spatial resolution are inferior to that of a monochromatic camera.
The essence of SRPIM idea can be explained by the concept of a non-spherical shape metallic nanoparticle (e.g. nanorod), which is a simple object with a polarization-dependent LSPR response. Its hypothetical reflection spectrum for linearly polarized light is shown in Fig.~\ref{fig:rodspec}. Resonance peaks appear at wavelengths $\lambda_{1}$ and $\lambda_{2}$ when the polarization axis is parallel to the minor and the major axes, respectively. In this case, we suppose that there are two nanorods with a separation distance less than the optical resolution, with their major axes set in different directions. When these nanorods are illuminated with a broad wavelength source or non-polarized light, the resulting optical images appear as spherical clusters due to diffraction. It is difficult to recognize two separate nanorods if the distance between them is less than the optical resolution. On the other hand, if illumination is performed with linearly polarized monochromatic light (for example$\lambda$ =$\lambda_{2}$), then the acquired image of the morphology should be different. When the polarization axis is set in the same direction as the major axis of the left side of the nanorods, only this side satisfies the resonance condition (Fig.~\ref{fig:spectrum} (b)). Therefore, a stronger reflection will be observed from this region and the morphology of the optical image is biased to the left side. If the polarization axis is changed to align with the direction of the major axis of the right side of the nanorods (Fig.~\ref{fig:spectrum} (c)), then the optical brightness peak will shift to this region and the morphology of the optical image will be biased to the right side. 
Based on this optical characteristic, it is possible to avoid the case where two grains are simultaneously bright, hence, the diffraction effect is circumvented. This is the principle of the proposed super-resolution approach. The barycenter shift of an optical image indicates the existence of optically unresolved nanoparticles and can be used to detect the track direction by tracking the barycenter movement.


\section{Nanometric tracking by super-resolution plasmon imaging microscopy}
\subsection{Optical microscopy and analytical method}
\begin{figure}[htbp]
 \begin{center}
   \includegraphics[width=70mm]{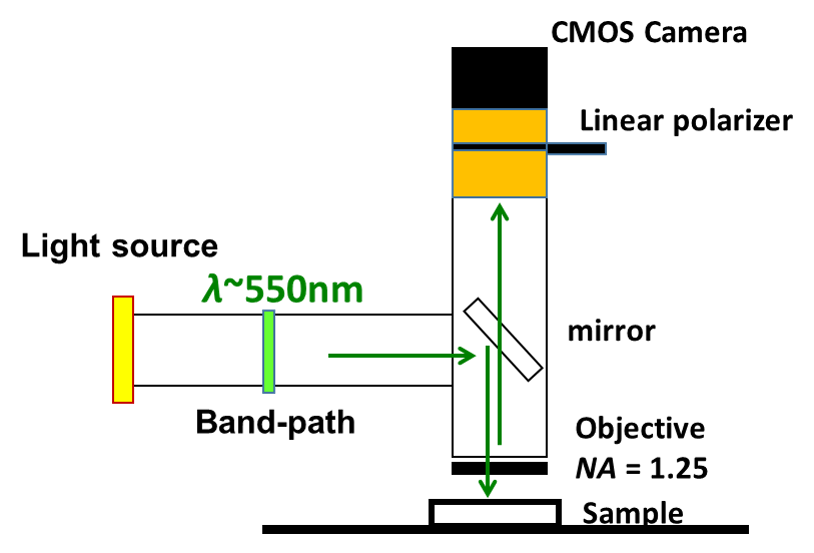}
 \caption{Schematics of the optical system based on an epi-illumination microscope for super-resolution plasmon imaging.} 
 \label{fig:PTS}
 \end{center}
\end{figure}

~~To demonstrate the concept described in Section. 2, we utilized an epi-illumination microscope equipped with an oil immersion objective lens (x100, NA=1.25) and a CMOS camera (Dalsa Falcon FA-21-1M120). The schematic of this setup is shown in Fig.~\ref{fig:PTS}. The resolution of image sensor was 1024 $\times$ 1024 pixel with 55 nm of the objective pixel size and the selected effective area of image sensor was 50 $\leq$ x, y $\leq$ 950 to avoid a coma aberration.
 To observe the polarization dependence caused by the LSPR, a linear polarizer was placed before the camera with a manually as rotating folder to identify the reflection of the linearly polarized light. To achieve monochromatic incident light, a bandpass filter ($\lambda$ = 550 nm, FWHM = 25 nm) was inserted after the light source. A mercury xenon lamp with a bright line at 550 nm was used. The optical spatial resolution was measured as 315 $\pm$ 15 nm using 60 nm silver spherical particles \cite{kgw}. A set of images were acquired using a rotating polarize, by rotation in increments of 10$^\circ$ between 0$^\circ$ and 180$^\circ$. The barycenter for each cluster found in the images was analyzed.\\

\subsection{Performance}
\subsubsection{Spatial resolution}
\begin{figure}[htbp]
 \begin{center}
   \includegraphics[width=140mm]{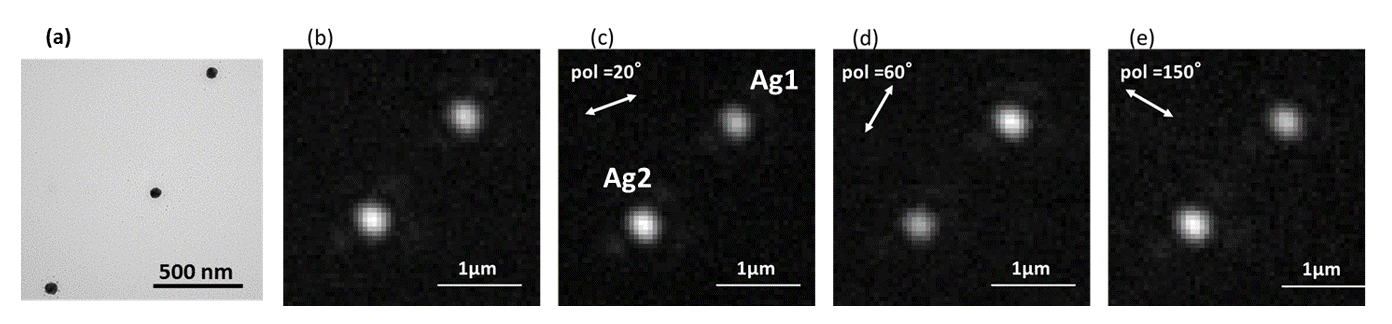}
 \caption{TEM image (a) and optical images of spherical nanoparticles.
Optical image (b) was acquired before setting the linear polarizer in the optical system, images (c),(d), and (e) are acquired with polarizer angle 20$^\circ$, 60$^\circ$, 150$^\circ$, respectively. The two particles change their brightness in relation to each other during rotation of the polarizer because their shapes are not perfectly spherical.} 
 \label{fig:fig7}
 \end{center}
\end{figure}

$\newline$
~~~Initially, the spatial resolution of the SRPIM method was measured as a position dispersion (i.e., positioning accuracy) of the barycenter shift of the object. Spherical 60 nm silver nanoparticles (SIGMA-ALDRICH Product number 730815) were used because no variation of the resonance and, therefore, no barycenter shift is expected after rotation of the polarizer for a spherical object. A Transmission Electron Microscope (TEM) image of the nanoparticle is shown in Fig.~\ref{fig:fig7}(a). 
The positioning accuracy acquired using this method can be regarded as the effective spatial resolution of the SRPIM for the given setup. The barycenter was determined from the “mean” of a Gaussian fit of the image brightness and position aberration caused by mechanical vibration during the rotation of the polarizer was corrected via image pattern matching. Since the fitting for the low image contrast event did not work right, we removed the events whose brightness difference between the maximum and background was smaller than 20 from the analysis targets. Optical images of silver nanoparticles are shown in Fig.~\ref{fig:fig7}. Two particles (Ag1 and Ag2) exist in the frame. The optical image (b) was acquired before setting the polarizer and (c), (d), and (e) were acquired with polarization axes of 20$^\circ$, 60$^\circ$, 150$^\circ$, respectively. Each polarization angle is identified as an arrow in Fig.~\ref{fig:fig7}. Since the shape of the nanoparticles is mostly spherical as confirmed by the TEM images, the shape represented in the images of the optical microscope does not indicate any difference for the different polarization axes. \\

\begin{figure}[htbp]
 \begin{center}
   \includegraphics[width=120mm]{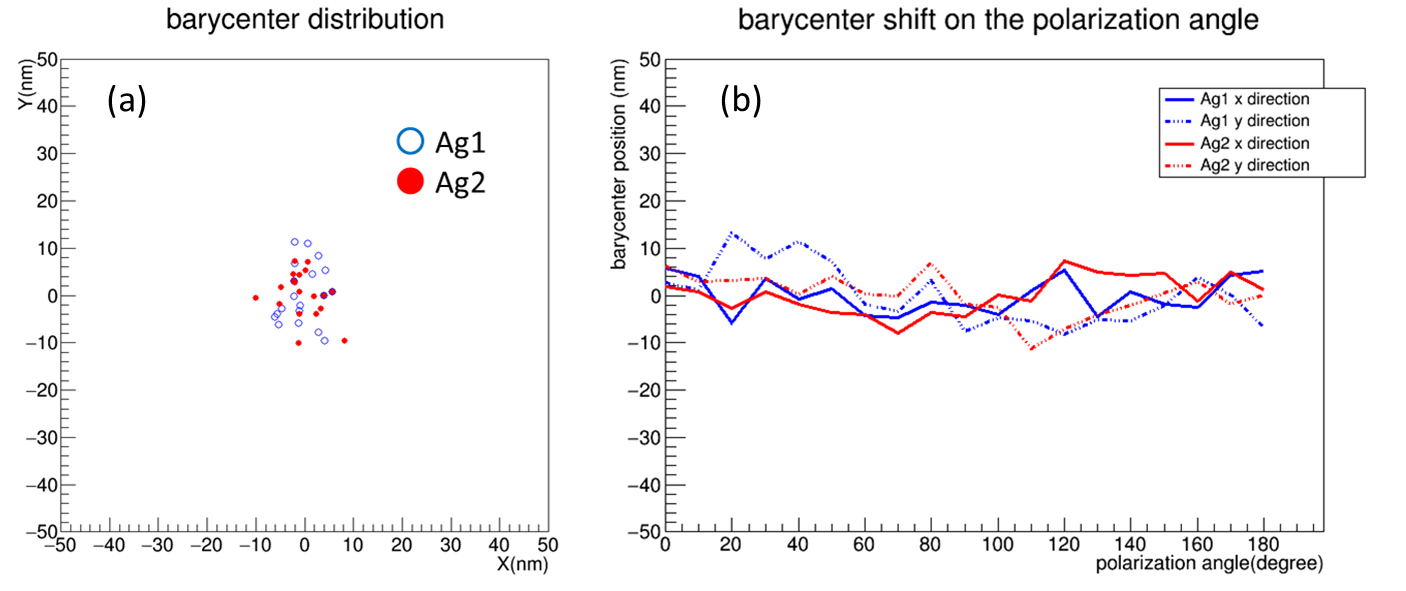}
 \caption{(a): Barycenter distribution of spherical nanoparticles for polarization analysis. A point is plotted as the mean position of each barycenter of Ag1 and Ag2 at a given polarization angle.
(b): Barycenter distribution along the polarization angle.The color represents a track ID and the line type indicates the X or Y direction} 
 \label{fig:fig8}
 \end{center}
\end{figure}

Fig~\ref{fig:fig8} (a) shows the barycenter distribution of Ag1 and Ag2 in the X -Y space of each image from their average barycenter measurement. The barycenter dispersion of both Ag1 and Ag2 were smaller than approximately 10 nm.
Fig.~\ref{fig:fig8} (b) shows the relation between the polarization angle and barycenter with respect to X or Y direction. The horizontal axis and the vertical axis indicate the polarization angle and the barycenter displacement with regards to the mean position for the X-direction (solid line) and the Y-direction (dashed line). The dispersions of barycenter originate from fluctuation statics, because the barycenter has no relation with polarization angle. The positioning accuracy can be determined from the standard deviation of the barycenter distributions. The accuracy is 5 $\pm$ 1 nm and it can be regarded as the spatial resolution in this method.\\

\subsubsection{Low velocity ion track detection}
\begin{figure}[htbp]
 \begin{center}
   \includegraphics[width=80mm]{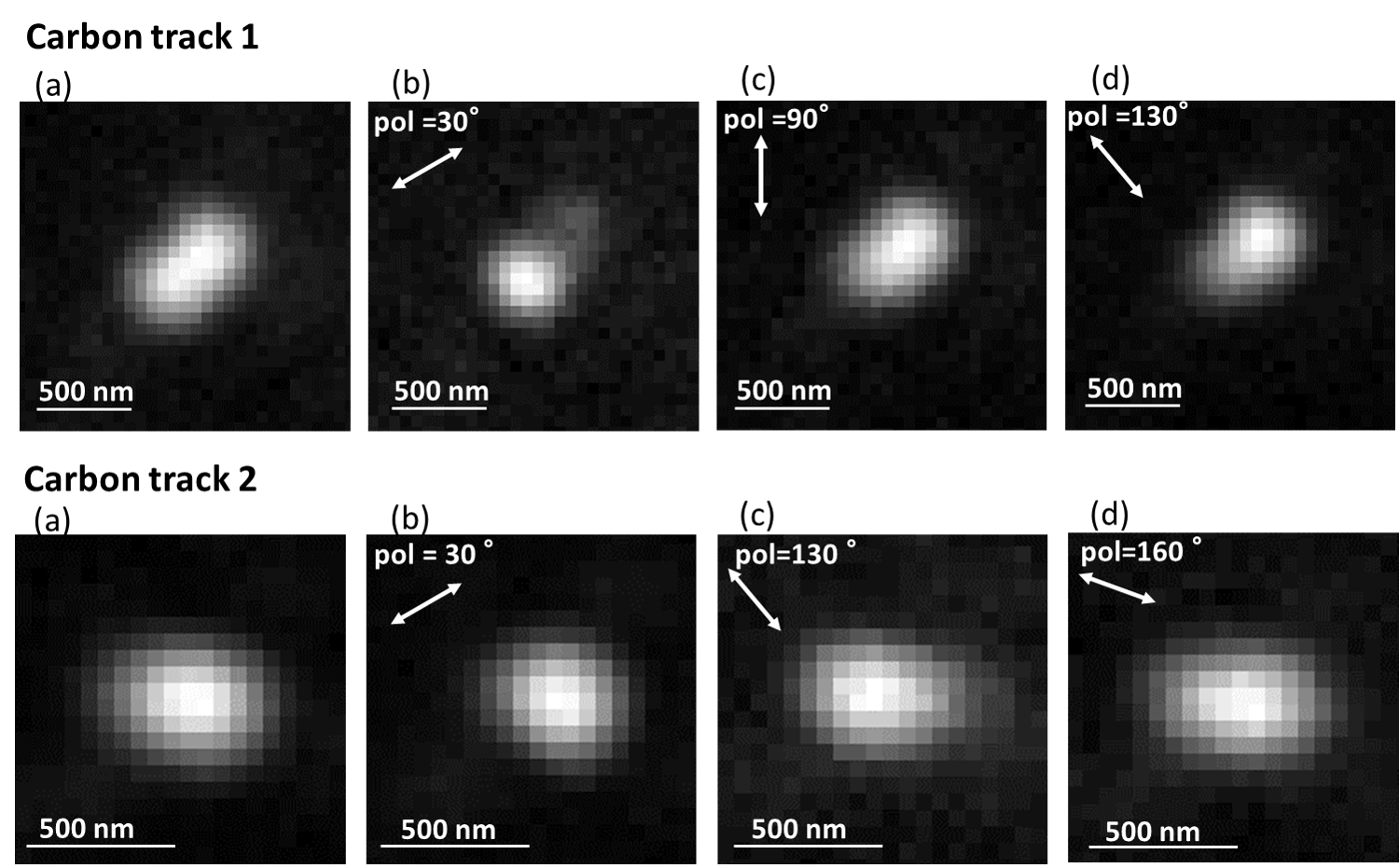}
 \caption{Optical images of 100 keV carbon track detected by NIT. Optical image (a) was acquired without setting the polarizer whereas images (b), (c), and (d) were acquired at different polarizer angles. } 
 \label{fig:fig9}
 \end{center}
\end{figure}

$\newline$
~~~Tracks of carbon ion with kinetic energy of 100 keV recorded in the NIT, as noted in section 2, was used. Those ions create tracks with a mean length of 270 nm in the NIT and the length is less than the optical resolution (315 $\pm$ 15 nm). They were therefore considered as evaluation objects to demonstrate the performance of the proposed technique. In Fig.~\ref{fig:fig9}, two characteristic carbon ion tracks are shown. Optical image (a) is acquired without a polarizer whereas optical images (b), (c), and (d) were acquired using different polarization angles as shown in Fig.~\ref{fig:fig7}. In the case of carbon track 1, it has a clear elliptical shape in image (a). However, the shape changes with the polarizer angle set to 30$^\circ$, 90$^\circ$, 130$^\circ$. It was observed that the upper right part region in image (b) with a lower intensity is brighter in image (d). A similar effect is also observed for carbon track 2. It appears to be nearly spherical in image (a), but a distinct change of the shape was observed after rotation of the polarization.\\ 

\begin{figure}[htbp]
 \begin{center}
   \includegraphics[width=120mm]{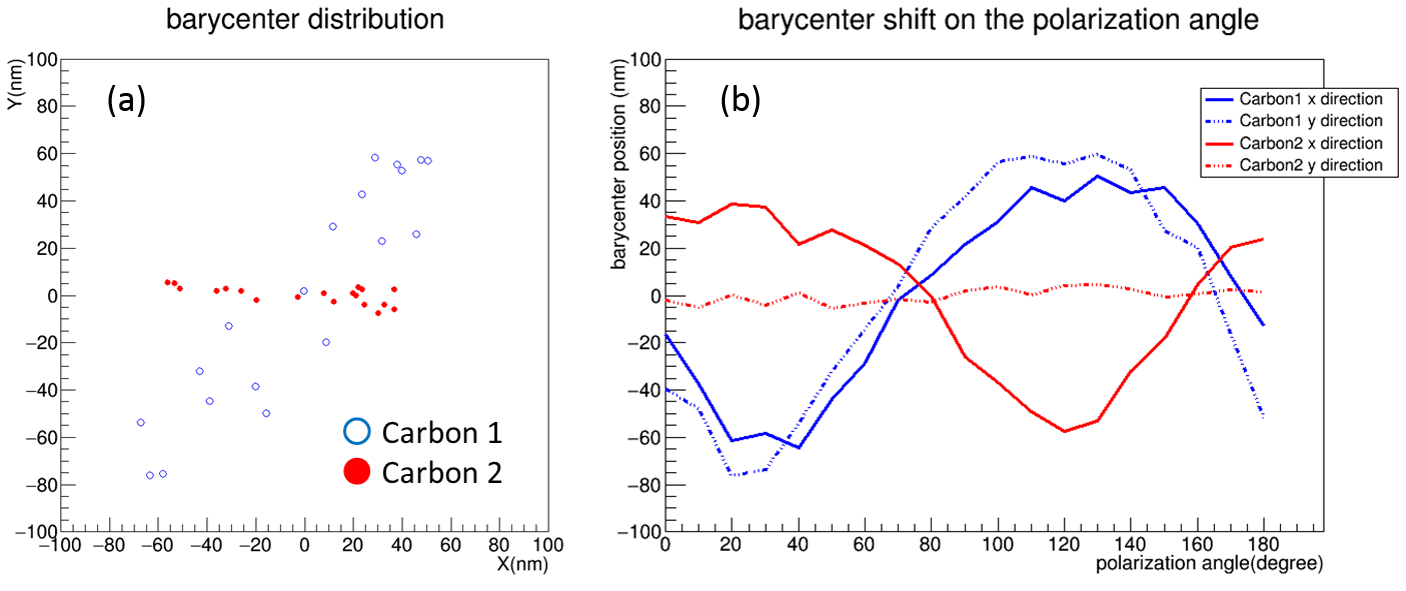}
 \caption{(a) shows the barycenter distribution of carbon track 1 and track 2 based on polarization analysis. 
(b) shows the relationship between barycenter position and polarizer angle.} 
 \label{fig:fig10}
 \end{center}
\end{figure}
\begin{figure}[htbp]
 \begin{center}
   \includegraphics[width=100mm]{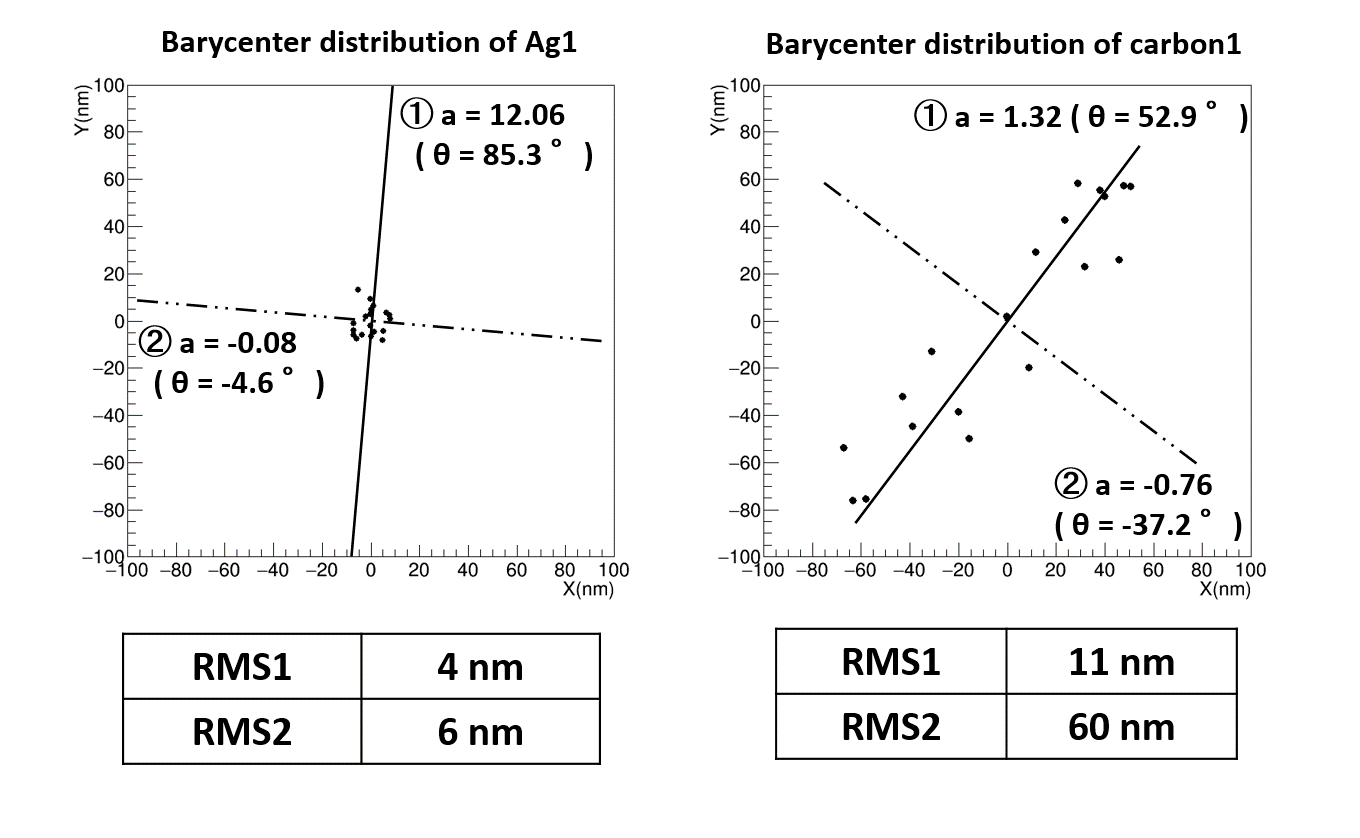}
 \caption{Result of a linear approximation method to discriminate a track with a single nanoparticle.} 
 \label{fig:fig11}
 \end{center}
\end{figure}

The barycenter position of carbon track 1 and track 2 are shown in Fig.~\ref{fig:fig10} (a) and the relationship between the barycenter position and the polarizer angle are shown in Fig.~\ref{fig:fig10} (b). In Fig.~\ref{fig:fig10} (a) and (b), it was determined that the barycenter moves linearly, and its modulation depends on the polarizer angle. The barycenter shift is much larger than the spatial resolution, which was defined based on the results obtained for the silver nanoparticles in the previous section. Barycenter shifts of length 170 nm and 100 nm were observed for carbon1 and carbon2, respectively.
The movement of the barycenter position should have a relation with the direction of the track-grains. As shown in Fig.~\ref{fig:fig11}, the track angle is obtained from the slope of the barycenter distribution. This slope can be parameterized by applying a linear approximation method. In this case, the approximate straight line (Line 1) is also used to determine whether an event is a track or not. The distance from each barycenter position to Line 1 is calculated and the root mean square of the distances (RMS1) was obtained. Then, another straight line (Line 2) perpendicular to Line 1 was used to obtain the root mean square of the distance from each barycenter position (RMS2) to the Line 2. Fig.~\ref{fig:fig11} shows the barycenter distribution with Line 1 and Line 2 whose slope parameters,  the gradient "a" and the angle "$\theta$," are included in the graph. The left graph represented the data of Ag1 and the right one is that of carbon1. Both RMS1 and RMS2 are indicated in the table below Fig.~\ref{fig:fig11}. In the case of Ag1, there is no anisotropy in the barycenter distribution, i.e., RMS1 and RMS2 are almost the same. On the other hand, Line 1 of carbon1 is fitted well along the distribution and RMS2 is notably larger than RMS1 as illustrated in Fig.~\ref{fig:fig11}. Also, it was determined that RMS1 of carbon 1 was significantly larger than the position accuracy. This might be due to filament structure of track-grains and track-grains that are not on the straight line. This result also shows that the SRPIM has a spatial resolution of several nanometers. 

\begin{figure}[htbp]
 \begin{center}
   \includegraphics[width=120mm]{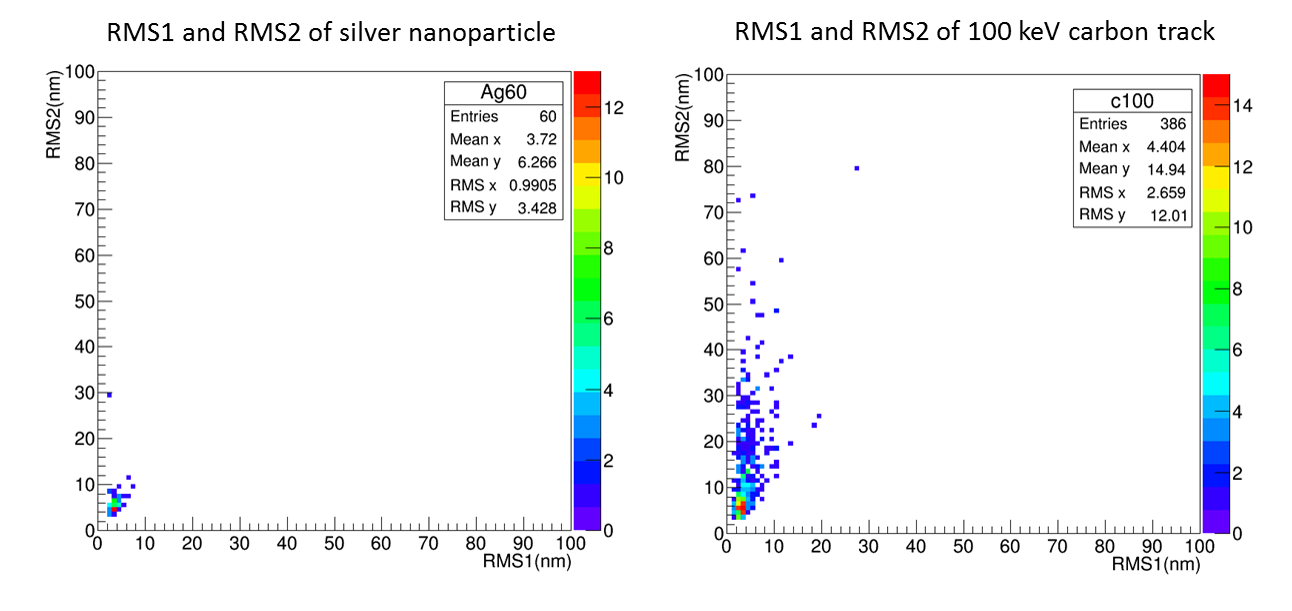}
 \caption{Relation between RMS1 and RMS2.The figure on left side shows the result of silver nanoparticles
and right figure shows the result of 100 keV carbon ion tracks.} 
 \label{fig:fig12}
 \end{center}
\end{figure}

In NIT device, there is a single silver nanograin called “fog“, a type of noise that is randomly generated from the AgBr crystal during several processes (e.g. image development process). RMS1 and RMS2 are useful selection parameters to discriminate a track from a fog by choosing events with large RMS2 in comparison to RMS1.  
The distributions of RMS1 and RMS2 for silver nanoparticles and 100keV carbon tracks are plotted in Fig.~\ref{fig:fig12}. The left graph shows the result of silver nanoparticles and it is determined that almost all events such that RMS1 and RMS2 are less than 10 nm. The only one event has the RMS2 around 30 nm and it is thought to be a cluster caused by the ununiformity in the dispersion of silver nanoparticle solution. On the other hand, there are many events with RMS2 $>$ RMS1 in the 100 keV carbon track data plotted on the right side in Fig.~\ref{fig:fig12}. 

\begin{figure}[htbp]
 \begin{center}
   \includegraphics[width=80mm]{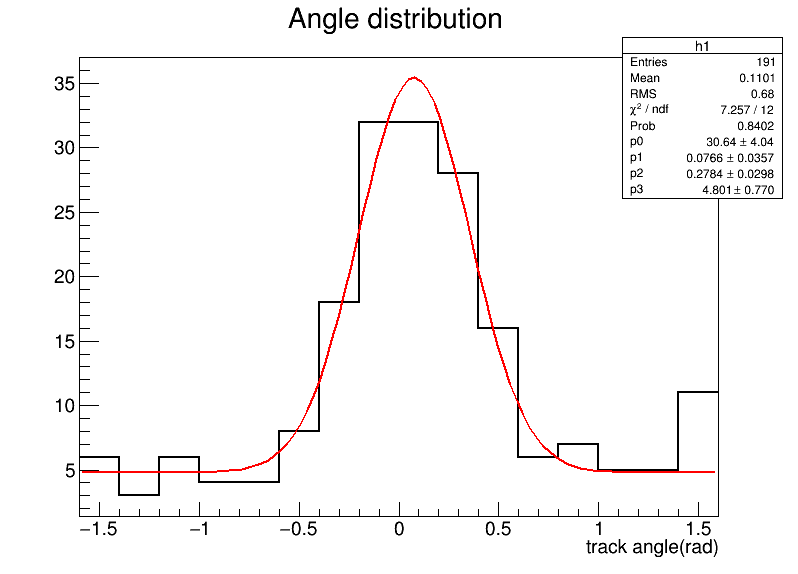}
 \caption{100 keV carbon ion track angle distribution obtained by the SRPIM.} 
 \label{fig:fig13}
 \end{center}
\end{figure}

Then, we set RMS2 $\geq$ 2$\times$RMS1 with RMS2 $\geq$ 10 as the threshold for track-like event selection.
This threshold corresponds to the situation when the movement of the barycenter is detected more than the dispersion due to the filament structure of single track-grain with two sigma significance.
 The track angle (i.e., Line1 slope) is plotted in Fig.~\ref{fig:fig13}. The distribution is well fitted by a Gaussian curve and the peak occurs at 0 rad., which is the known carbon ion beam direction. The standard deviation of the angular distribution 17$\pm$ 2$^\circ$ is consistent with the simulation which takes into account the effect of multiple scattering \cite{SRIM}. 

When the above threshold was applied as the event selection, the event rate of 100 keV carbon ions with mean track length of 270 nm was approximately 49 $\pm$ 4 \% . We hereby demonstrate that the SRPIM have the spatial resolution of 5 $\pm$ 1 nm and it is capable of identifying tracks and detecting track angles beyond the optical resolution limit (i.e., 315 $\pm$ 15 nm).\\

\section{Conclusion}
~~ We demonstrated a technique for super-resolution plasmonic imaging (SRPIM) to detect tracks with shorter length than the optical resolution of a diffraction-limited optical microscopy system. SRPIM is a new technique which combines super fine-grained NIT emulsions and LSPR. The track detection capability of this method was evaluated using 100 keV carbon ion tracks consisting of silver nanograins with filament structures and 60 nm spherical silver spherical nanoparticles. 
The barycenter was defined from the peak brightness of the distribution using a rotating polarizer and it was determined that the position accuracy to identify the silver nanoparticles position was 5$\pm$1 nm. This value corresponds to the spatial resolution of the SPRIM technique. It was determined that the 100 keV carbon tracks had a larger barycenter movement than the spatial resolution, and a linear fitting method was developed to identify the track based on angular information. 
We showed that SRPIM could identify carbon track beyond the theoretical optical resolution. As a next step, an automatic rotator will be used to change the polarization angle and online-image processing will be developed to achieve the efficiency of scanning. The optical response of the track-grains strongly depends on their structure. However, the shape and size of track-grains can be controlled by appropriate selection of the solution for image development. For example, a complicated filament structure is obtained by physical development process and a spherical silver grain structure with a gold coating on the surface is obtained when gold deposit development is used \cite{kubota}. As such, we are considering a change of the track-grain structure for active LSPR response.

\vskip2pc

\section*{Acknowledgment}    

 This work was supported by a Grant-in-Aid for JSPS Research Fellows Grant Numbers 16J03974 and JSPS KAKENHI Grant Numbers JP15H05446,JP 18H03699,JP26104005.
 The electron microscopy was performed at the Division for Medical Research Engineering, Nagoya University Graduate School of Medicine.  A part of this work was conducted at the Nagoya University Nanofabrication Platform, supported by the "Nanotechnology Platform Program" of the Ministry of Education, Culture, Sports, Science and Technology (MEXT), Japan.

\end{document}